\documentclass[letter,twocolumn]{jpsj3}
\usepackage{graphicx}
\usepackage{dcolumn}
\usepackage{bm}

\begin{document}

\title{Generic First Order Orientation Transition of Vortex Lattices in Type II Superconductors}

\author{Kenta M. Suzuki, Kenji Inoue, Predrag Miranovi\' c$^1$, Masanori Ichioka,\\
and Kazushige Machida}
%\affiliation{Department of Physics, Okayama University, % for APS
%Okayama 700-8530, Japan,
%$^1$Department of Physics, University of Montenegro,
%81000 Podgorica, Montenegro}
%\date{\today}                    % for APS
\inst{Department of Physics, Okayama University,
Okayama 700-8530\\
$^1$Department of Physics, University of Montenegro,
81000 Podgorica, Montenegro}
%\recdate{\today}

%\begin{abstract}    % for APS
\abst{%
First order transition of vortex lattices (VL) observed in various superconductors
with four-fold symmetry is explained microscopically by quasi-classical Eilenberger theory combined
with non-local London theory.
This transition is intrinsic 
%We identify two microscopic main sources of 
in the generic successive VL phase transition 
%including this first order one 
due to either gap or Fermi velocity anisotropies. 
This is also suggested by the electronic states around vortices. 
Ultimate origin of this phenomenon is attributed to some what hidden frustrations of a 
spontaneous symmetry broken hexagonal VL on the
underlying four-fold crystalline symmetry. 
%\end{abstract}        % for APS
}

%\pacs{74.25.Qt, 74.25.Op, 74.25.Jb, 74.20.Rp}
\kword{vortex lattice transition, anisotropic superconductors, Eilenberger theory,
non-local London theory}
%74. Superconductivity (for superconducting devices, see 85.25.?j) 
%74.20.Rp Pairing symmetries (other than s-wave)  
%74.25.Jb Electronic structure  
%74.25.Op Mixed states, critical fields, and surface sheaths  
%74.25.Qt Vortex lattices, flux pinning, flux creep  

\maketitle

%%%%%%%%% Introduction %%%%%%%%%%%%%%%%%%%%%%%%%%%%%%%%%%%%%%%%%%%%%%%%%%%%%%%%%%%%%%%%%%%%%%%%%%%%%%%%%%%%%%

Morphology of vortex lattices (VL) in the Shubnikov (mixed) state of type II superconductors is not
completely understood microscopically\cite{weber,brandt}. 
Thorough understanding of the VL symmetries 
and its orientation with respect to the underlying crystalline lattice are important both 
from the fundamental physics point of view 
because through those studies we can obtain the information on pairing symmetry (see below).
It is also important from technological application of a superconductor,
such as vortex pinning mechanism which ultimately determines critical current density. 

The first neutron diffraction experiment\cite{neutron} was conducted for Nb to observe a periodic array of the Abrikosov 
vortices.
%\cite{nb}. 
Concurrently since then various experimental methods 
have been developed, such as Bitter decoration of magnetic fine particles, $\mu$SR, NMR or scanning 
tunneling microscope (STM)\cite{weber,brandt}. The former ones are to observe spatial distribution of the 
magnetic field in VL and the last one is to observe the electronic structure of VL.
The variety of experiments compile a large amount of information on VL morphology.
Thus we need to synthesize many fragmented pieces of information which has been not done yet.  
Here we are going to investigate a mystery known for a quite some time and to demonstrate that the
present methodology is powerful enough which might be useful for systematic studies of Shubnikov state in 
type II superconductors.

\begin{figure}
\includegraphics[width=7cm]{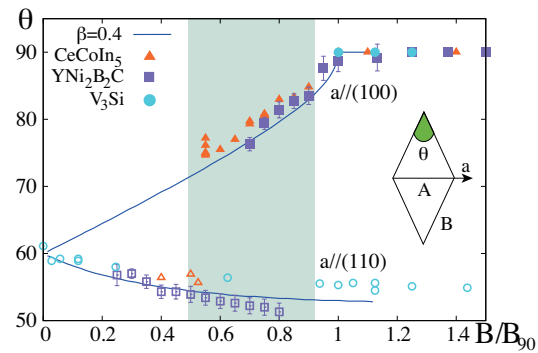}

%\vspace{-0.5cm} 
\caption{(Color online) 
Field dependence of open angle $\theta$ for V$_3$Si ($\bigcirc$)\cite{v3si-yathiraj,v3si-sosolik},
YNi$_2$B$_2$C($\Box$)\cite{y-levett,y-dewhurst,y-paul} and CeCoIn$_5$($\bigtriangleup$)\cite{eskildsen}.
The open and filled symbols, respectively, belong to orientation 
${\bm a} \parallel (110)$ and ${\bm a} \parallel (100)$.
The shaded band indicates first order transition region for these materials.
Lines are theoretical curves by non-local London theory with
$\beta=0.4$ for guide of eyes. 
Inset shows a unit cell of VL with definitions of
open angle $\theta$, orientation direction ${\bm a}$,
and bonds $A$ and $B$ of VL. 
}
\label{fig:theta1}
\end{figure}

As shown in Fig. \ref{fig:theta1} 
where the open angle $\theta$ of a unit cell is plotted as a function of magnetic  
field $B$ applied to the four-fold symmetric axis (001) 
in either  cubic crystal (V$_3$Si\cite{v3si-yathiraj,v3si-sosolik})
or tetragonal crystals (YNi$_2$B$_2$C\cite{y-levett,y-dewhurst,y-paul}, CeCoIn$_5$\cite{eskildsen}), 
the observed VL symmetries change
in common, yet they belong to completely different superconducting classes (conventional or unconventional):
Namely starting with the regular triangle lattice at the lower critical field $B_{c1}$,
the VL transforms into a general hexagonal lattice with $\theta\neq 60^{\circ}$,
keeping the orientation ${\bm a}$ the same: parallel to (110). 
This direction is defined as (110) here, so that ${\bm a} \parallel (110)$. 
The inset of Fig. \ref{fig:theta1} shows the definition of $\theta$ and the orientation ${\bm a}$.  
At a certain field $B_{\rm 1st}$ (shaded area in Fig. \ref{fig:theta1})
whose values depends in external conditions, such as cooling rate, etc.,
the orientation ${\bm a}$ suddenly changes by 
angle $45^{\circ}$ to ${\bm a}\parallel (100)$ at a first order phase transition. 
Upon further increasing $B$, $\theta$ continuously increases and finally becomes $\theta=90^{\circ}$ at $B_{90}$.
This lock-in transition at $B=B_{90}$ is of second order. 
Those successive symmetry changes in VL are common for those compounds. 
Other A-15 compounds Nb$_3$Sn\cite{furukawa} or TmNi$_2$B$_2$C\cite{morten} exhibit a similar trend.

We explain this phenomenon by synthesizing two theoretical frameworks:
phenomenological non-local London theory~\cite{kogan,kogan96,affleck} 
and microscopic Eilerberger theory\cite{eilenberger,ichioka,nakai}. 
%Before going into detailed calculations,
%we give an intuitive physical picture to 
%this and put our problem in a broader perspective.
The purpose of this study is to examine the universal behaviors of first order transition in the  
VL morphology under changing the degree of anisotropies coming from two sources: the Fermi velocity
anisotropy and superconducting gap anisotropy. 
In this analysis, we evaluate the validity of non-local London theory, 
comparing with quantitative estimates by Eilenberger theory. 
Further, we discuss intrinsic reasons why the first order transition of VL orientation occurs, 
in viewpoints from frustrations and electronic states around vortices.   
Microscopic calculation is necessary to discuss the electronic states. 

As we will see below, frustrations are a crucial key concept to understand this phenomenon.
Antiferromagnetic spins on a triangular lattice is a well known example
of frustration. Here a triangular VL is also a driving force due to hidden frustration
which is more subtle than the spin case,
leading to successive phase transition. 
Hexagonal symmetry is incompatible with four-fold symmetry of underlying crystal lattice,  
because all bonds between vortices are not satisfied simultaneously to
lower bonding energies.

Quasi-classical Eilenberger theory
\cite{eilenberger,ichioka,nakai} is valid for $k_F\xi \gg 1$ ($k_F$: the Fermi wave 
number and $\xi$: the coherence length) a condition met in
the superconductors of interest here. 
Eilenberger equations read as 
\begin{eqnarray} && 
\left(\omega_n+{\bm v}(\theta)\cdot({\bm \Pi}+i{\bm A})\right) f
=\Delta({\bm r},\theta) g,
\\ && 
\left(\omega_n-{\bm v}(\theta)\cdot({\bm \Pi} -i{\bf A})\right) f^\dagger
=\Delta^*({\bm r},\theta) g
\qquad 
\end{eqnarray} 
with $g^2+ff^\dagger=1$ 
for quasi-classical Green's functions 
$g(\omega_n,{\bm r},\theta)$, $f(\omega_n,{\bm r},\theta)$, 
and $f^\dagger(\omega_n,{\bm r},\theta)$.  
%Here $\bm \Pi=\bm\nabla+(2\pi i/\Phi_0)\bm A$ is gauge invariant gradient, 
%$\bm A$ is vector-potential and $\Phi_0$ is flux quantum;
In our formulation, we use Eilenberger unit 
where length, magnetic field and temperature are, respectively, scaled by 
$R_0=\hbar v_{\rm F0}/2 \pi k_{\rm B} T_{\rm c}$,  
$B= \phi_0/2 \pi R_0^2$, and transition temperature $T_{\rm c}$ 
with flux quantum $\phi_0$. 
Matsubara frequency $\omega_n=T(2n+1)$ with integer $n$.
%Normalization condition for quasi-classical Green's functions $g,f,f^\dagger$; $g^2+ff^\dagger=1$.
The pairing interaction is assumed separable 
$V(\theta,\theta')=V_0 \phi(\theta)\phi(\theta')$ 
so that gap function
is $\Delta(\bm r,\theta)=\Psi(\bm r)\phi(\theta)$.
Since we consider two-dimensional case with cylindrical Fermi surface, 
we set the normalized Fermi velocity ${\bm v}(\theta)\equiv{\bm v}_{\rm F}/v_{\rm F0}$ as 
${\bm v}(\theta)=v(\theta) (\cos\theta,\sin\theta)$. 
When we discuss the  Fermi velocity anisotropy,  we model it as
$v(\theta)=(1+\beta\cos{4\theta})/\sqrt{1-\beta^2}$ and $\phi(\theta)=1$,
which is called $\beta$ model~\cite{nakai}. 
When we discuss the gap anisotropy, $v(\theta)=1$ and 
%$\phi(\theta)^2=\phi({\pi\over 8})^2(1-\alpha\cos{4\theta})$
$\phi(\theta)^2=1-\alpha\cos{4\theta}$
as $\alpha$-model (anisotropic $s$-wave), or 
$\phi(\theta)=\sqrt{2} \cos 2\theta$ as $d$-wave pairing. 
Here $\theta$ is the polar angle relative to $(100)$ axis. 
%Constant $\phi({\pi\over 8})^2=1/(1+(1-\sqrt{1-\beta^2})\alpha/\beta)$ is chosen to
%assure the same $T_c$ and density of states $N(0)$ for any
%value of anisotropy parameters $\alpha$ and $\beta$.

The selfconsistent equations for the gap function $\Psi(\bm r)$ and
vector-potential ${\bm A}({\bm r})$ are
\begin{eqnarray} && 
\Psi(\bm r)=V_0 N_0 \ 2T \sum\limits_{\omega_n>0}
\left\langle \phi^\ast(\theta) f \right\rangle 
\label{selfgap}
\\ && 
{\bm\nabla}\times{\bm\nabla}\times{\bm A}({\bm r})=
-\frac{2T}{\kappa^2} 
\sum\limits_{\omega_n>0}{\rm Im\;} \left\langle{\bm v}g \right\rangle.
\end{eqnarray}
with $V_0 N_0=\ln T + 2T \sum_{\omega_n>0} \omega_n^{-1}$ 
and Ginzburg Landau parameter $\kappa$.  
For average over Fermi surface,
$\langle \ldots \rangle =\int_0^{2 \pi}(\ldots) d\theta/2 \pi v(\theta)$  
with extra factor $1/v(\theta)$
coming from angle-resolved density of states $N(\theta)=N_0/v(\theta)$
on Fermi surface.
The self-consistent
solution yields a complete set of the physical quantities: 
the spatial profiles of the order parameter $\Psi(\bm r)$ and the magnetic field $B(\bm r)$.
The local density of sates (LDOS) for electronic states is calculated by 
$N({\bm r},E)=N_0 {\rm Re} 
\langle g(\omega_n, {\bm r},\theta)|_{{\rm i}\omega_n \rightarrow E + {\rm i}0} \rangle$.  
The free energy density is given by
\begin{equation*}
F= \kappa^2 \overline{  B^2(\bm r) } 
 -T \sum\limits_{\omega_n>0}
\overline{ \left\langle 
\dfrac{1-g}{1+g}(\Delta^* f+\Delta f^\dagger)
\right\rangle } .
\end{equation*} 
Here, $\overline{(\ldots)} $ is a spatial average within a unit cell of VL.
Free energy should be minimized with 
respect to the VL symmetry and its orientation relative to the 
crystallographic axes. 
In previous studies by Eilenberger theory, 
the first order transition of the VL orientation was not evaluated at low $T$, 
while only the transition from triangular to square VL was 
discussed~\cite{ichioka,nakai}. 
In principle, we can investigate the whole space spanned by $(B,T)$. 
But in practice it is not easy
to exhaust the parameters in order to seek the desired physics.
Thus our calculations are backed up by the non-local London theory. 
%,which is valid when $B\sim B_{c1}$ and low $T$.

%%%% Formulation of non-local London theory %%%%%%%%%%%%%%%%%%%%%%%%%%%%%%%%%%%%%%%%%%%%%%%%%

The non-local London theory is powerful and handy for $B_{c1}\lesssim B$ at low $T$. 
The nonlocal relation between current 
$\bm j$ and vector potential $\bm A$ in Fourier space is derived as  
$\frac{4 \pi}{c} j_l({\bm q})
=i[{\bm q}\times{\bm A}]_l=-\kappa^{-2} \sum_m Q_{lm}({\bm q})A_m({\bm q})$ 
($l,m=x,y$) from the Eilenberger theory~\cite{kogan96}. 
Since we use the Eilenberger unit here, 
the penetration depth $\lambda$ is changed to $\kappa$ in the length unit $R_0$ ($\sim\xi$). 
The kernel is 
\begin{equation}
Q_{lm}({\bm q})
=2T \sum_{\omega_n>0} \left\langle 
\frac{|\Delta(\theta)|^2 v_l v_m }
{\beta_n(\beta_n^2+({\bm v}\cdot{\bm q})^2)} 
\right\rangle
\end{equation} 
with
$\beta^2_n=\omega_n^2+\Delta^2(\theta)$, and $\Delta(\theta)$ is a uniform solution 
of the gap function. 
%, and $\gamma({\bf q})={1\over 2}{\bf v}\cdot{\bf q}$.
In previous studies by the non-local London theory for the first order transition in borocarbides,  
higher order terms than $q^4$ was neglected~\cite{kogan}.   
Here, we do not expand $Q_{lm}$ by $q$ so that we can include all order contributions of $q$ in $Q_{lm}$~\cite{affleck}.
The corresponding London free energy is given by
%$$F_L={1\over {8\pi}}\Sigma_{\bf q}[{\bf B}({\bf q})^2+({\bf q}\times {\bf B}
%({\bf q})){\bf Q}^{-1}({\bf q})({\bf q}\times {\bf B}({\bf q}))].$$
\begin{eqnarray}&&
F_L=\sum_{\bm q} 
{\rm e}^{-\frac{1}{2} \xi^2 q^2}
[1+\kappa^2 \{(Q^{-1})_{yy} q_x^2
\nonumber \\ && \hspace{2cm}
 + (Q^{-1})_{xx} q_y^2 -2(Q^{-1})_{xy} q_x q_y\} ]^{-1}, 
\qquad
\end{eqnarray} 
where $Q^{-1}$ is the inverse matrix of $Q_{lm}$ depending on ${\bm q}$. 
The non-local London theory is valid near $B_{c1}$ at lower $T$ where the vortex core contribution is 
approximated as a cutoff parameter $\xi$ mimicking the finite core size effect. 
To explore wide ranges of $(B,T)$ on a firm basis, 
we need to carefully check the validity of non-local London theory, using Eilenberger theory. 
We report our results for $\kappa=89$ and $T=0.2 T_{\rm c}$ in both theories. The results do not 
depend on the $\kappa$ value unless $\kappa$ is approaching $1/\sqrt 2$.

%%%%%%%%%%%%%%% study for beta-model %%%%%%%%%%%%%%%%%%%%%%%%%%%%%%%%%%%%%%%%%%%%%

\begin{figure}
\includegraphics[width=6cm]{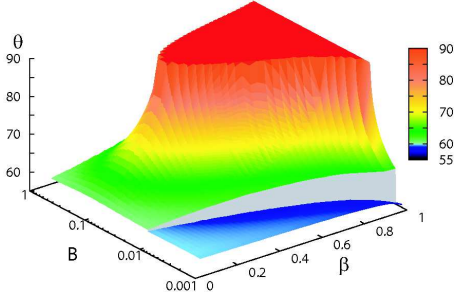}

%\vspace{-0.5cm} 
\caption{(Color online) 
Stereographic view of the open angle $\theta$ as functions of 
$B$ and $\beta$ in the $\beta$-model by non-local London theory. 
%$T/T_c=0.2$.
First order transitions from $\theta < 60^\circ$ 
to $\theta > 60^\circ$ with  45$^{\circ}$ rotation of VL orientation 
are seen for any values of $\beta$.
}
\label{fig:beta-theta}
\end{figure}

First, we study the $\beta$-model, i.e., anisotropy of Fermi velocity. 
Figure \ref{fig:beta-theta} shows a stereographic view of the open angle $\theta$
as functions of $B$ and $\beta$.
It is seen that 
(1) irrespective of the  $\beta$ values the first order transition exists
seen as a jump of $\theta$ in Fig. \ref{fig:beta-theta}. 
(2) The jump of the first order transition becomes large as $\beta$ increases.
(3) $\beta>\beta_{\rm cr}=0.38$ ($\beta<\beta_{\rm cr}$) at $T/T_c=0.2$, the square lattice is (never) realized.
(4) This square lattice becomes ultimately unstable for higher fields.
Items (3) and (4) were also confirmed by Eilenberger calculation 
[see Fig. 2(a) in Ref. \citen{nakai}].
Note in passing that in CeCoIn$_{5}$ the square lattice changes into a
hexagonal lattice at a higher field, which reminds us the similarity,
but we believe that it is caused by other effect, such as the Pauli paramagnetic
effect\cite{hiasa,kenta}.

\begin{figure}
\includegraphics[width=8.5cm]{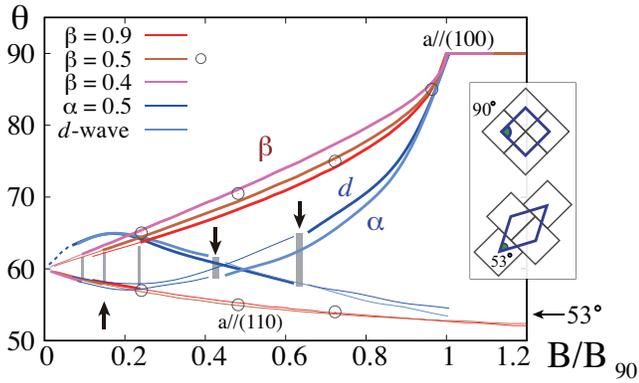}

%\vspace{-0.5cm} 
\caption{(Color online) 
Open angle $\theta$ as a function of $B/B_{90}$ for two orientations 
in the $\beta$-model ($\beta=0.4, 0.5, 0.9$) by non-Local London, 
and in the $\alpha$-model ($\alpha=0.5$) and $d$-wave model by Eilenberger theory. 
Circles show $\theta$ obtained by Eilenberger theory in the $\beta$-model with $\beta=0.5$. 
%$T/T_c=0.2$.
Arrows indicate $B_{\rm 1st}$. 
Inset shows the closed packed square tiles and the associated hexagonal unit cell, 
indicating $\theta=53^{\circ}$ (lower inset) and $\theta=90^\circ$ (upper inset) 
when $B \rightarrow B_{90}$. 
%Note that the theoretical minimum open
%angle $\theta_{min}$ is limited to $53^{\circ}$
}
\label{fig:theta2}
\end{figure}

When plotted as a function of $B/B_{90}$ by respective lock-in field $B_{90}$ 
as shown in Fig. \ref{fig:theta2}, 
the $\theta$-behaviors are similar for any $\beta$ $(> \beta_{\rm cr})$, 
while $B_{\rm 1st}$ changes depending on $\beta$.  
These universal behaviors well reproduce the experimental data as shown in Fig. \ref{fig:theta1}.  
We also plot some points of $\theta$ obtained by Eilenberger theory for $\beta=0.5$ (see the circle symbols in Fig. 3), 
which shows similar behavior as in the non-local London theory. 
Therefore, the non-local London theory can be applied reliably to superconductors  
whose anisotropy comes from the Fermi velocity. 

%%%%%%%%%%%%%%% study for alpha-model & d-wave %%%%%%%%%%%%%%%%%%%%%%%%%%%%%%%%%%%%%%%%%%%%%

Next, we study the case of anisotropic pairing gap by the $\alpha$-model and $d$-wave pairing. 
Figure \ref{fig:theta2} also shows $\theta$-behavior in these cases obtained by Eilenberger theory.  
Even in the gap anisotropy, we find the first order transition of orientation, 
where the $\theta$-dependence is not monotonic at low $B$ 
and $B_{\rm 1st}$ is higher compared to the $\beta$-model. 
Thus, independent of the sources of anisotropy (Fermi velocity or gap), 
there is always the first order orientational transition 
in the similar successive VL transition.
Starting with $\theta=60^{\circ}$ at $B_{\rm c1}$, 
the ${\bm a}$-direction of VL coincides with the gap minimum ($\alpha$-model) 
or with the Fermi velocity minimum ($\beta$-model).
This orientation is changed via a first order transition 
to the orientation where the ${\bm a}$-direction is
rotated by 45$^{\circ}$. 
However, in the gap anisotropy case, the reentrant transition from square to hexagonal VL 
at high fields does not occur. 
We note that in these anisotropic gap cases,  at low $T$ 
the non-local theory does not work to reproduce the $\theta$-behaviors 
of Eilenberger calculation~\cite{affleck}.

It is also interesting to notice  simple geometry
that the open angle $\theta_{\rm min}=2\tan ^{-1} ({1\over 2})=53^{\circ}$
when square tiles with same size are closely packed as shown in lower inset of Fig. \ref{fig:theta2}.
%The smaller $\theta$ makes those square tiles overlap, 
%leading to further loosing the condensation energy.
In the VL orientation ${\bm a}\parallel (110)$, as seen from Fig. \ref{fig:theta2}, 
the theoretical minimum of $\theta$ indicates $\theta_{\rm min}\sim 53^{\circ}$ for both anisotropy cases, 
which is roughly obeyed by the experimental data shown in Fig. \ref{fig:theta1}.
In the other orientation ${\bm a}\parallel (100)$, $\theta \rightarrow 90^\circ$ 
at $B \rightarrow B_{90}$. 
This is also understandable by the packing of square tiles in different way,  
as shown in upper inset of Fig. \ref{fig:theta2}.

%%%%%%%%%%%%%%% LDOS %%%%%%%%%%%%%%%%%%%%%%%%%%%%%%%%%%%%%%%%%%%%%

\begin{figure}
\includegraphics[width=8cm]{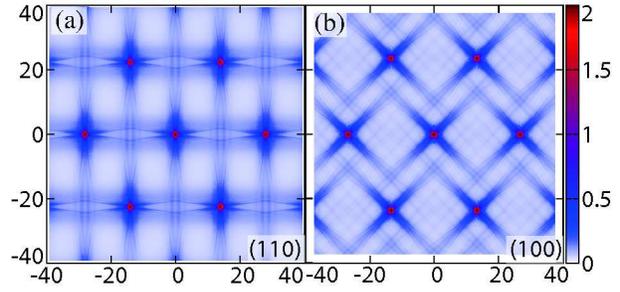}

%\vspace{-1.0cm} (a) \hspace{7cm} (b)

%\vspace{0.5cm}

\caption{(Color online) 
Zero energy LDOS $N({\bm r},E=0)/N_0$ for hexagonal VL with $\theta \sim 60^\circ$ at $B/B_{90}=0.07$ 
in $d_{x^2-y^2}$-wave pairing, where the gap minimum is along (110) direction. 
(a) Stable orientation ${\bm a}\parallel (110)$. 
(b) Unstable orientation ${\bm a}\parallel (100)$.  
Horizontal axis is parallel to ${\bm a}$-direction. 
}
\label{fig:LDOS1}
\end{figure}

\begin{figure}
\includegraphics[width=8.5cm]{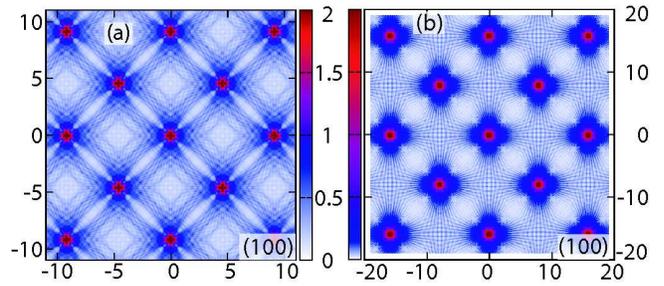}

%\vspace{-1.0cm} (a) \hspace{7cm} (b)

%\vspace{0.5cm}

\caption{(Color online) 
Zero energy LDOS $N({\bm r},E=0)/N_0$ for the square VL 
with ${\bm a}\parallel (100)$ at $B/B_{90}=1.2$ 
in the $d_{x^2-y^2}$-wave case (a)  and the $\beta$-model with $\beta=0.5$ (b).
}
\label{fig:LDOS2}
\end{figure}

It is instructive to examine the electronic structures of various VL's to
understand the origins of the successive transition.
First, we discuss stable orientation of hexagonal VL at low $B$ 
in the $d_{x^2-y^2}$-wave pairing. 
In Fig. \ref{fig:LDOS1} 
we display the typical LDOS at zero energy (i.e., Fermi level) for two orientations. 
%We compare two orientations in the hexagonal lattice 
%in Figs. \ref{fig:LDOS1}(a) and \ref{fig:LDOS1}(b). 
%The former (latter) is energetically stable (unstable) orientation.
%As mentioned before, in the hexagonal VL the $a$-axis coincides with the nodal direction 
%in $d$-wave case, 
In stable orientation ${\bm a}\parallel(110)$ in (a), 
the zero energy LDOS is well connected between nearest neighbor (NN) vortices along ${\bm a}$, 
and between next NN vortices. 
Here ${\bm a}$ is parallel to node direction of $d$-wave pairing, connected by the A-type bond in inset of Fig. 1. 
These interconnections effectively lower the kinetic energy of quasi-particles,
leading to a stable VL symmetry and orientation.
In contrast, in the VL of unstable orientation in (b), 
the interconnections are not well organized, clearly
demonstrating it less favorable orientation and VL symmetry energetically.
We note that, even in the stable hexagonal VL [Fig. \ref{fig:LDOS1}(a)],  
four B type bonds defined in inset of Fig. 1 among the six NN vortex bonds are not favorable directions for interconnections. 
Thus those are frustrated, which ultimately leads to further successive VL transitions
in higher fields.

%In the unit cell of a regular triangular VL there are three NN bonds which are classified into  A-bond and two B-bonds 
%shown in the inset of Fig. \ref{fig:theta1}. 
%The A-bond is satisfactory when the $a$-axis is on the gap minimum
%direction, but the remaining two B-bonds are off the gap minimum direction.
%The other possible orientation is to orient  the $a$-axis to the gap maximum direction.
%In this orientation all three bonds in a unit cell are more or less unsatisfactory, 
%resulting in the unstable orientation.
%In this respect there is not completely satisfactory  hexagonal VL under four-fold symmetry.
%The hexagonal VL's are intrinsically frustrated.
%Thus even in the VL with the stable orientation, the VL deforms by adjusting $\theta$
%upon increasing $B$ from $B_{c1}$ where the vortex spacing comes closer. 

The concept of interconnections between NN vortices via zero energy LDOS 
continues to be useful for square VL at high fields.
In $d$-wave pairing as shown in Fig. \ref{fig:LDOS2}(a),
interconnections are highly well organized to stabilize it. 
In particular not only all NN vortex connections are tightly bound, but also second, 
third neighbor vortices are also connected.
This high field stable VL configuration can not be continuously transformed from 
low field stable hexagonal VL in Fig. \ref{fig:LDOS1}(a)just  by changing $\theta$. 
Therefore there must exist a first order phase transition in an intermediate field region 
to rotate the orientation by $45^\circ$. 
This intuitive explanation is basically correct for a superconductor with the four-fold anisotropy  
coming from the Fermi velocity anisotropy.

In the $\beta$-model shown in Fig. \ref{fig:LDOS2}(b) in the stable square VL, 
the zero energy LDOS extends broadly to neighbor vortices.  
%and there are only a few lines to direct towards the NN vortices.
This weak connection between vortices ultimately leads to the instability for this square VL in the
$\beta$-model towards a hexagonal VL in higher fields as shown in Fig. \ref{fig:beta-theta}. 
This may be one of the reasons for the reentrance phenomenon in the $\beta$-model. 
Note that in the $\alpha$-model
there is no reentrance up to $B_{c2}$ and the square VL is most stable at high fields.

%This suggests that the $\beta$-model is more subtle than the $\alpha$-model.
%In fact, in the $\alpha$-model the successive transition is now easily understandable:
%From these analysis of stable orientation, we understand that the first order transition of orientation 
%Starting with the regular triangular lattice with $\theta=60^{\circ}$ and the $a$-axis
%being parallel to the gap minimum, VL tends to directly go to the square VL by ``increasing''
%$\theta$, which leads to the ``wrong'' square VL (see Fig. \ref{fig:theta2}). 
%This movement is interrupted by a first order orientational transition because the tending
%square VL is ``wrong'' square VL.

%In contrast, the $\beta$-model is more subtle in this sequence because 
%we must care about the condensation energy explicitly.
%It is known in this model that the effective core radius is anisotropic:
%It is longer (shorter) towards the Fermi velocity maximum (minimum) direction\cite{ichioka}.
%Near $B_{c1}$ in the regular triangle VL the overlapping of the condensation energy landscape occurs along the 
%NN direction. To relax this overlapping which losses the condensation energy in order to realize the closed packing, 
%$\theta$ tends to ``decrease'' upon increasing $B$ 
%from $B_{c1}$. This explains the different behaviors of $\theta$ near $B_{c1}$
%as shown in Fig. \ref{fig:theta2} where $\theta$ increases (decreases) from $\theta=60^{\circ}$ 
%for the $\alpha$ ($\beta$)-model.

%%%%%%%%%%%% Discussions and summary

Let us examine the experimental data in light of the present calculations.
V$_3$Si\cite{v3si-yathiraj,v3si-sosolik} whose data are shown by circle symbols in Fig. 1 and Nb$_3$Sn\cite{furukawa} are cubic crystals with A-15 structure and known to 
be an $s$-wave superconductor with an isotropic gap.
The main four-fold anisotropy comes from the Fermi velocity, 
thus an example of the $\beta$-model.
YNi$_2$B$_2$C\cite{y-levett,y-dewhurst,y-paul} (the squares in Fig.1) has tetragonal crystal and  is  known 
to be $d_{xy}$-like gap symmetry\cite{nishida}, thus an
example of the $\alpha$-model albeit the Fermi velocity anisotropy may be also working.
In fact, the successive transition precisely follows our calculation, ending up with the square VL whose NN
is directed to (100) \cite{y-levett,y-dewhurst,y-paul}. 
There is no indication for the reentrance transition up to $B_{c2}$.
CeCoIn$_5$\cite{eskildsen} (the triangles in Fig.1) has a $d_{x^2-y^2}$ gap symmetry. 
The successive transition, 
including first order, ends up with the square VL which orients along 
(110) as expected. Towards  $B_{c2}$ it exhibits the reentrant transitions to
the hexagonal VL with the same orientation near $B_{c1}$.
Similar reentrant VL transition is also found in TmNi$_2$B$_2$C\cite{morten}.
This intriguing reentrance phenomenon, which is not covered here,
belonging to a future problem because we need to consider the Pauli paramagnetic
effect\cite{hiasa,kenta}.

In conclusion, 
universal behaviors of the first order transition 
in the VL morphology from hexagonal to square lattice have been studied  
by the non-local London theory and Eilenberger theory. 
We find that non-local London theory is accurate  for the superconductor 
whose anisotropy comes from the Fermi velocity, 
while it is uncontrollable for
the gap anisotropy case, which is covered by Eilenberger theory.
%Two sources of the anisotropy are the ultimate origin for the whole phenomena
%of interest here.
We have explained the successive VL transition with first order one
observed in various four-fold symmetric superconductors.
Since the origin of this phenomenon is due to either gap anisotropy
or Fermi velocity anisotropy, which are present in any superconductors,
it is desirable to carefully perform experiments of neutron diffraction, $\mu$SR, NMR or STM
to observe this generic phenomenon.
In particular it is interesting to see it in high $T_c$ cuprates, Sr$_2$RuO$_4$
and TmNi$_2$B$_2$C for $H//(001)$, which are known to have four-fold gap
anisotropy, yet so far this phenomenon has not been reported
although square VL's are observed.

\end{document}